\newcommand{\hst}{\textit{HST}}
\newcommand{\spitzer}{\textit{Spitzer}}

\newcommand{\ks}{\hbox{$K_s$}}

\newcommand{\ha}{\hbox{H$\alpha$}}
\newcommand{\lsim}{\lesssim}
\newcommand{\gsim}{\gtrsim}

\newcommand{\mcal}{\hbox{$\mathcal{M}$}}

\newcommand{\etal}{et al.}

\newcommand{\msol}{\hbox{$\mathcal{M}_\odot$}}

\newcommand{\lsol}{\hbox{$L_\odot$}}

\newcommand{\infinity}{\hbox{$\infty$}}

\newcommand{\jmk}{\hbox{$J - K_s$}}

\newcommand{\ujy}{\hbox{$\mu$Jy}}
\newcommand{\lir}{\hbox{$L_{\mathrm{IR}}$}}
\newcommand{\mone}{\hbox{$[3.6\mu\mathrm{m}]$}}
\newcommand{\mtwo}{\hbox{$[4.5\mu\mathrm{m}]$}}
\newcommand{\mthree}{\hbox{$[5.8\mu\mathrm{m}]$}}
\newcommand{\mfour}{\hbox{$[8.0\mu\mathrm{m}]$}}
\newcommand{\micron}{\hbox{$\mu$m}}
\newcommand{\arcmin}{\hbox{$^\prime$}}

\documentclass[ptsize 1]{ppmo}                  
\RequirePackage{amssymb}%
\usepackage{fancyhdr}
\usepackage{graphicx}                   
\input{epsf.sty}                        
\input{psfig.sty}                       

\begin{document}

\pagestyle{fancy}

   \title{A Spitzer View of Massive Galaxies at $z\sim 1-3$}
   \volnopage{Vol.0 (2005) No.0, 000--000}      
   \setcounter{page}{1}          
   \baselineskip=5mm             

   \author{Casey Papovich\mailto{papovich@as.arizona.edu}
   \and the GOODS and MIPS GTO teams
      }

   \institute{Steward Observatory, 933 N.\ Cherry
Ave., Tucson, Arizona, 85721, USA \\
             \email{papovich@as.arizona.edu}
         }

   \date{Received~~2005 month day; accepted~~2006~~month day}

   \abstract{ I discuss constraints on star--formation and AGN activity
   in massive galaxies at $z$$\sim$1--3 using observations from the
   \spitzer\ Space Telescope at 3--24~\micron.   In particular I focus
   on a sample of distant red galaxies (DRGs) with \jmk$>$2.3 in the
   southern Great Observatories Origins Deep Survey (GOODS--S) field.
   Based on their ACS (0.4--1~\micron), ISAAC (1--2.2~\micron), and
   IRAC (3--8~\micron) photometry, the DRGs have typical stellar
   masses $\mcal$$\gsim$$10^{11}$~\msol.  Interestingly, the majority
   ($\gsim$50\%) of these objects have 24~\micron\ detections, with
   $f_\nu(24\micron)$$\geq$50~\ujy.   If attributed to star formation,
   then this implies star--formation rates (SFRs) of
   $\simeq$100--1000~\msol\ yr$^{-1}$.    Thus, massive galaxies at
   $z$$\sim$1.5--3 have specific  SFRs equal to or exceeding the
   global average value at that epoch.  In contrast, galaxies with
   $\mcal$$\geq$$10^{11}$~\msol\ at $z$$\sim$0.3--0.75 have specific
   SFRs less than the global average, and more than an order of
   magnitude lower than that at $z$$\sim$1.5--3.  Thus, the bulk of
   assembly of massive galaxies is largely complete by $z$$\sim$1.5.
   At the same time, based on the X--ray luminosities and near-IR
   colors, as many as 25\% of the massive galaxies at $z$$\gsim$ 1.5
   host AGN, implying that the growth of supermassive black holes
   coincides with massive--galaxy assembly.   Lastly, the analysis of
   high--redshift galaxies depends on bolometric corrections between
   the observed \spitzer\ 24~\micron\ data and total IR luminosity.  I
   review some of the sources of the (significant) uncertainties on
   these corrections, and discuss improvements for the future.
 }

   \authorrunning{C.~Papovich}            
   \titlerunning{Spitzer View of Massive Galaxies at $z\sim 1-3$}  

\setlength\baselineskip {5mm}             

 \maketitle

%
%
\section{Introduction}           
Most ($\sim$50\%) of the stellar mass in galaxies today formed during
the short time between $z$$\sim$3 and 1 (e.g., Dickinson et al.\ 2003,
Rudnick et al.\ 2003).   Although much of this stellar mass density resides in
massive galaxies, which appear at epochs prior to $z$$\sim$1--2
(e.g., Bell et al.\ 2004, McCarthy 2004), it is still
unclear when and where the stars in these galaxies formed.  It may be
that galaxies ``downsize'', forming most of their stars in their current
configuration at early cosmological times, with lower mass galaxies
continuing to form stars to the present epoch (e.g., Bauer et al.\ 2005,
Juneau et al.\ 2005).  Alternatively, stars may form predominantly in
low--mass galaxies at high redshifts, which then merge
over time to form large, massive galaxies at more
recent times (Bauer et al.\ 1998, Kauffmann \& Charlot 1998, Cimatti
et al.\ 2002).

Understanding the formation and evolution of massive galaxies has been
challenged by difficulties in constructing complete samples of
star--forming and massive galaxies at $z$$\gsim$ 1.  Massive galaxies
at $z$$\lsim$1 exist on a fairly prominent red sequence (e.g., Blanton
et al.\ 2003, Bell et al.\ 2004, Faber et al.\ 2005), and are largely
devoid of ongoing star formation, evolve passively, and contain up to
half of the stellar--mass density.   However, current
hierarchical models predict colors for massive galaxies at $z$$\sim$0
that are too blue compared to observations (e.g., Somerville, Primack,
\& Faber 2001, Dav\'e et al.\ 2005).  Some recent models suppress star
formation at late times in massive galaxies by truncating star
formation at some mass threshold, or by using feedback from strong AGN
(e.g., Granato et al.\ 2001, Dav\'e et al.\ 2005, Croton et al.\
2006).   Most of the massive galaxies at $z$$\lsim$1 have red colors
with formation epochs $z_f$$\gsim$ 2.   The morphologies of
the most optically luminous (and most massive) galaxies transforms
from ``normal'' early--type galaxies at $z$$\sim$ 1 to irregular
systems at $z$$\sim$ 2--3 (e.g., Papovich et
al.\ 2005).  Therefore, we need to study the properties of massive
galaxies at these earlier epochs.

In these proceedings, I discuss recent \spitzer\ observations at
3--24~\micron\ of massive galaxies at $z$$\sim$1.5--3 in the southern
Great Observatories Origins Deep (GOODS--S) field.   The \spitzer\
IRAC and MIPS observations provide constraints on the star--formation
and AGN activity in massive galaxies at these epochs.  I discuss
implications the analyses of these data have for  the stellar--mass
assembly rates and formation epochs of massive galaxies.  I also
review some the uncertainties on these bolometric corrections between
the observed \spitzer\ 24~\micron\ data and total IR luminosities in
$z$$\sim$ 1.5--3 galaxies and prospects for future improvements.

\section{Stellar Masses and Star Formation in High--$z$ Massive Galaxies}

GOODS is a multiwavelength survey of two 10\arcmin$\times$15\arcmin\
fields, one in the northern \textit{Hubble} Deep Field, and one in the
southern \textit{Chandra} Deep Field.  The GOODS datasets include
(along with other things) \hst/ACS and VLT/ISAAC imaging (Giavalisco
et al.\ 2004), and recent \spitzer\ imaging.  I make use of these data
for the work described here, as well as data from
\spitzer/MIPS 24~\micron\ in this field from time allocated to
the MIPS GTOs (e.g., Papovich et al.\ 2004).

In these proceedings, I primarily focus on so--called distant red
galaxies (DRGs) selected with \jmk$>$2.3~mag (see Franx et al.\ 2003).
This color criterion identifies both galaxies at $z$$\sim$2--3.5
whose light is dominated by a passively evolving stellar population
older than $\sim$250~Myr (i.e., with a strong Balmer/4000~\AA\ break
between the $J$ and \ks--bands), and also star--forming galaxies at
these redshifts whose light is heavily reddened by dust
(F\"orster--Schreiber et al.\ 2004, Labb\'e et al.\ 2005, Papovich et
al.\ 2006).  For the GOODS--S data, the
\jmk$>$2.3~mag selection is approximately complete to stellar mass
$\mcal$$\geq$$10^{11}$~\msol\ for passively evolving galaxies, and we
find 153 DRGs to $\ks$$\leq$23.2~mag, spanning 0.8$\leq$$z$$\leq$3.7
with $\langle z\rangle $$\simeq$2.2 (see Papovich et al.\ 2006).

\begin{figure}[h]
  \begin{minipage}[th]{0.5\linewidth}
  \centering
  \includegraphics[width=0.95\linewidth]{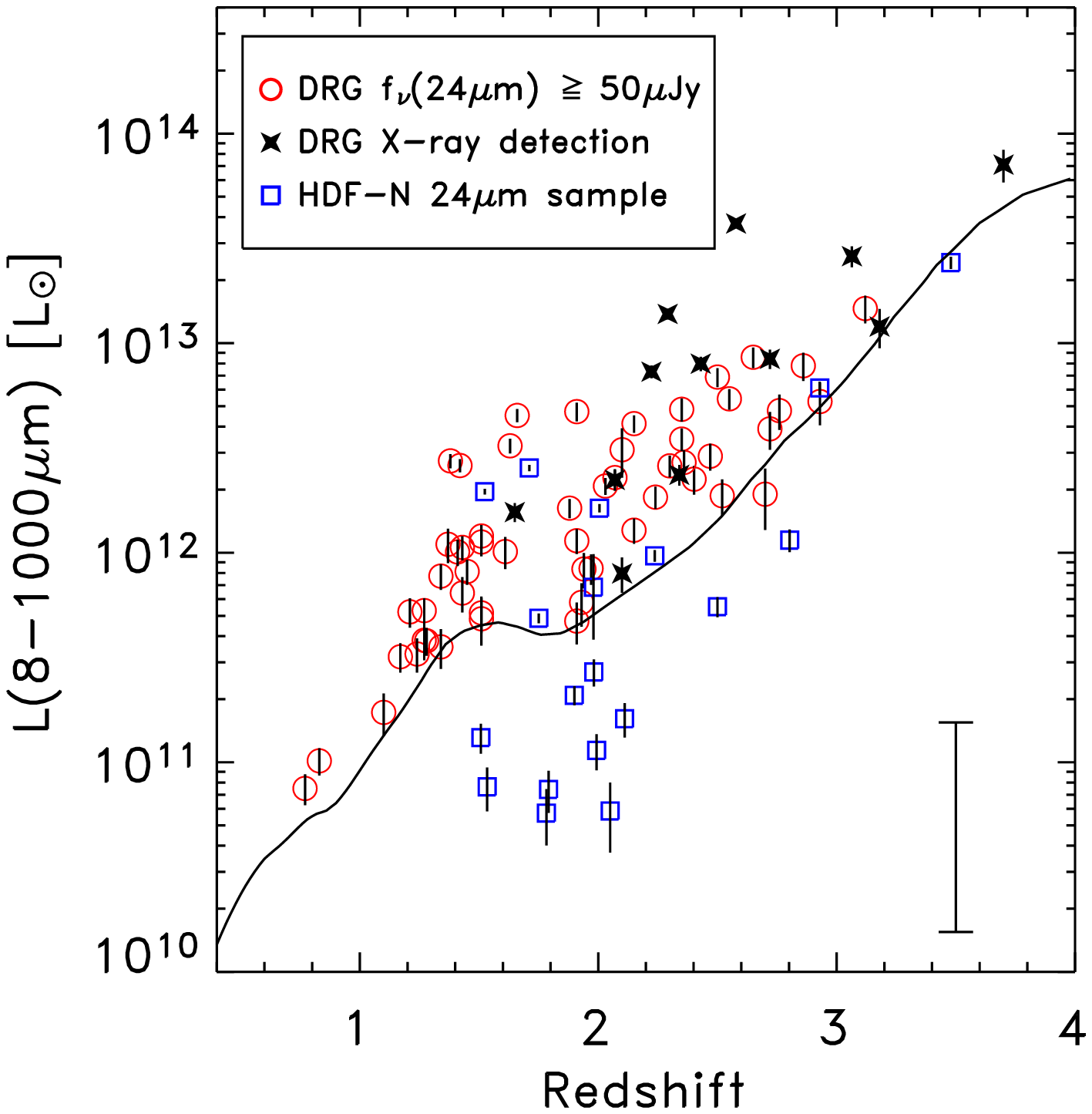}
\vspace{-18pt}
   \caption{Total IR luminosities, $\lir$$\equiv$$L(8-1000)$, of
   galaxies inferred from their observed 24~\micron\ emission.  Red
   circles show the values for the DRGs; black stars show those
   objects with X--ray detections.  Blue squares show objects from the
   HDF--N data (see Papovich et al.\ 2006).  The solid line denotes
   the 24~\micron\ 50\% completeness limit.   The inset error bar
   shows the estimated systematic error, $\approx$0.5~dex.}
   \end{minipage}%
\begin{minipage}[th]{0.5\textwidth} \centering
   \includegraphics[width=0.95\linewidth]{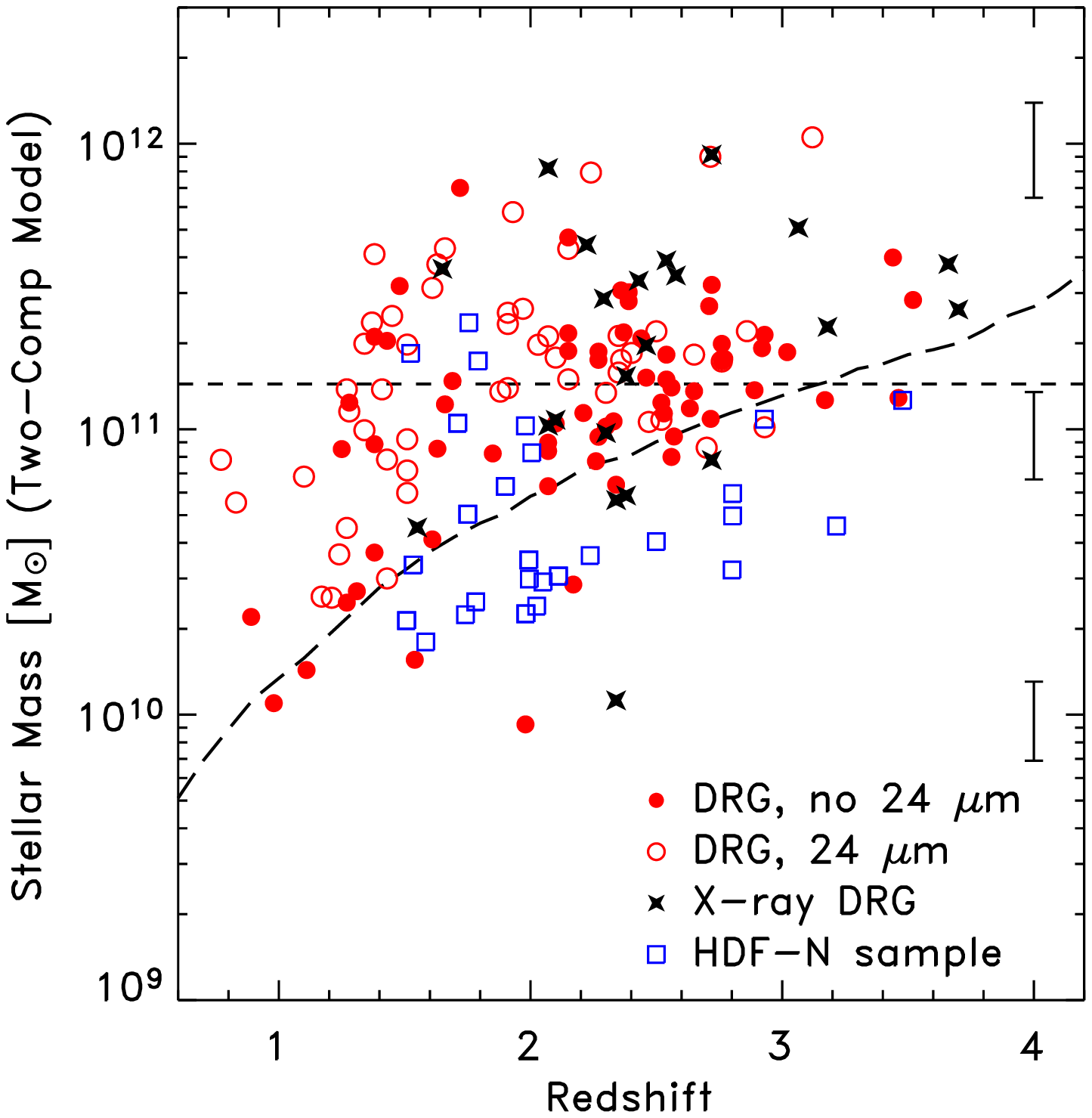}
\vspace{-18pt}
   \caption{Stellar masses of galaxies inferred by fitting
   two--component models to the galaxies' rest--frame UV--to--near-IR
   data.  The inset bars show the mean errors as a function of
   mass. The short--dashed line shows the characteristic present--day
   stellar mass (Cole et al.\ 2001); the long--dashed line shows the
   stellar mass limit for a passively evolving stellar population formed at
   $z$$\sim$$\infinity$ with $\ks$=23.2~mag. } \end{minipage}
\vspace{-20pt}
 \end{figure}

More than 50\% of the DRGs have 24~\micron\ detections with
$f_\nu(24\micron)$$\geq$50~\ujy.   Daddi et al.\ (2005) find a similar
24~\micron--detection rate for massive galaxies at
1.5$\lsim$$z$$\lsim$2.5 selected via their $BzK$ colors.
Interestingly, this implies that the majority of massive galaxies at
$z$$\sim$2 emit strongly in the thermal IR ---
\textit{they are either actively forming stars, supermassive blackholes, or
both at this epoch}.  The 24~\micron\ emission at $z$$\sim$1.5--3
probes the mid--IR ($\sim$5--10~\micron), which broadly correlates
with the total IR, $\lir$$\equiv$$L(8-1000\micron)$ (Chary \& Elbaz
2001).  Figure~1 shows the inferred
\lir\ for the DRGs using the Dale et al.\ (2002) models to convert the
observed mid--IR to total IR luminosity.  
There is inherent uncertainty in this conversion, which I discuss
in \S~3. The 24~\micron\ flux densities for the $z$$\sim$1.5--3 DRGs yield
$\lir$$\approx$$10^{11.5-13}$~\lsol, which if attributed to
star--formation corresponds to SFRs of $\approx$100--1000~\msol\
yr$^{-1}$ (e.g., Kennicutt 1998).  

Nearly all of the DRGs are detected in the deep \spitzer/IRAC data,
implying they have substantial stellar masses.  In Papovich et al.\
(2006), we modeled the DRG stellar populations by comparing their ACS, ISAAC,
and IRAC \mone\ \mtwo\  photometry to a suite of stellar--population synthesis
models (Bru\-zu\-al \& Char\-lot 2003),  varying the age,
star--formation history, and dust content.  We use the model
stellar-mass--to--light ratios to estimate the galaxies' stellar mass.
We first allow for star--formation histories with the SFR
parameterized as a decaying exponential with an $e$--folding time,
$\tau$, ranging from short $\tau$'s (burst of star--formation) to long
$\tau$'s (constant star--formation).  We also use models with a
two--component star--formation history characterized by a passively
evolving stellar population formed in a ``burst'' at
$z_\mathrm{form}$=$\infty$, summed with the
exponentially--decaying--SFR model.   Although the modeling loosely
constrains the ages, dust content, and star--formation histories of
the DRGs, it provides relatively robust estimates of the galaxies'
stellar masses (see also F\"orster--Schreiber et al.\ 2004).   Typical
uncertainties for the stellar masses for the full DRG sample are
0.1--0.3~dex. Figure~2 shows the stellar masses inferred for the DRGs
by fitting the two--component models.  

\begin{figure}[th]
\vbox{  \begin{minipage}[t]{0.5\linewidth}
  \centering
  \includegraphics[width=0.95\linewidth]{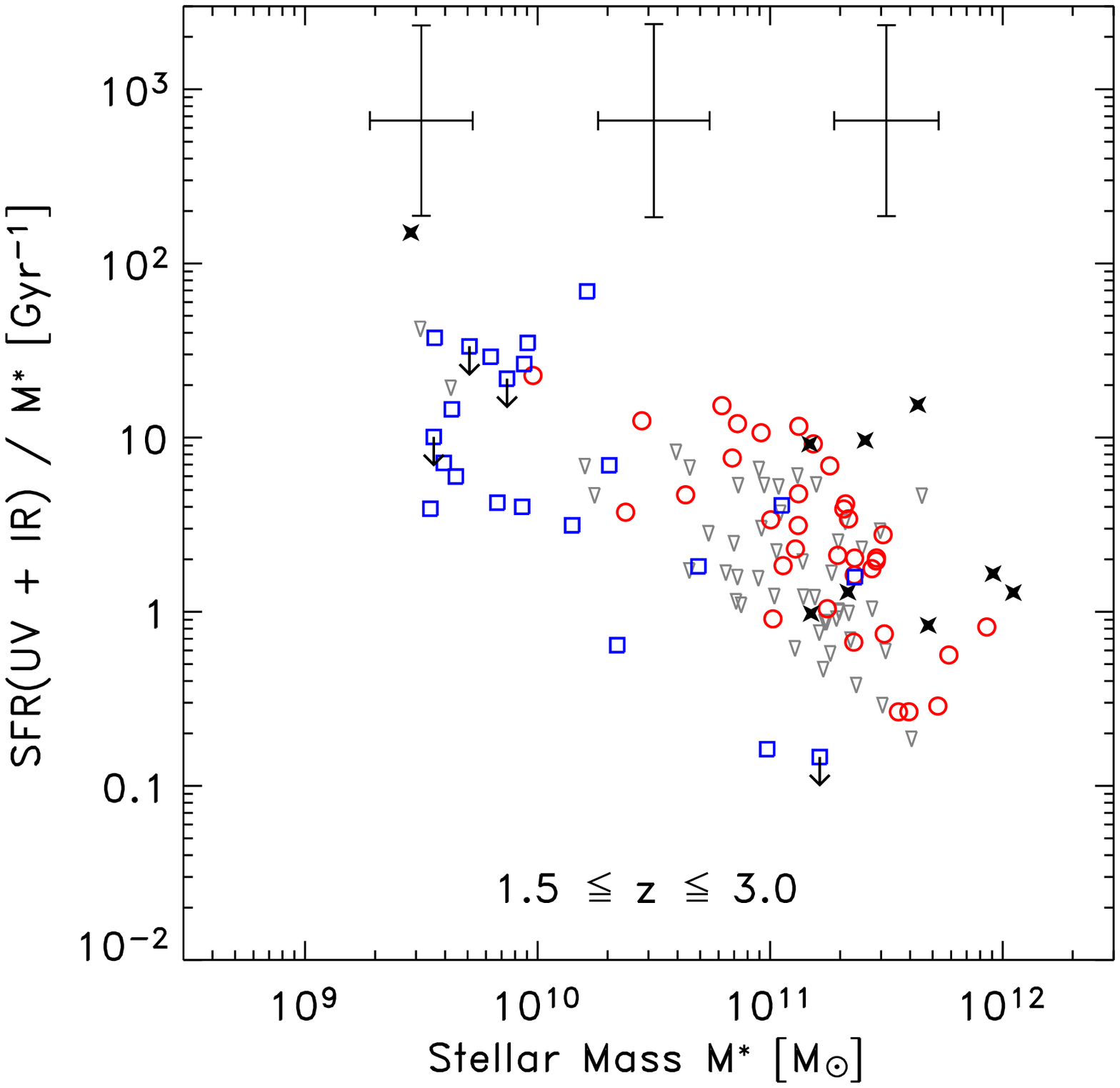}
   \label{fig:specsfrmass}
  \end{minipage}%
  \begin{minipage}[t]{0.5\textwidth}
  \centering
 \includegraphics[width=0.95\linewidth]{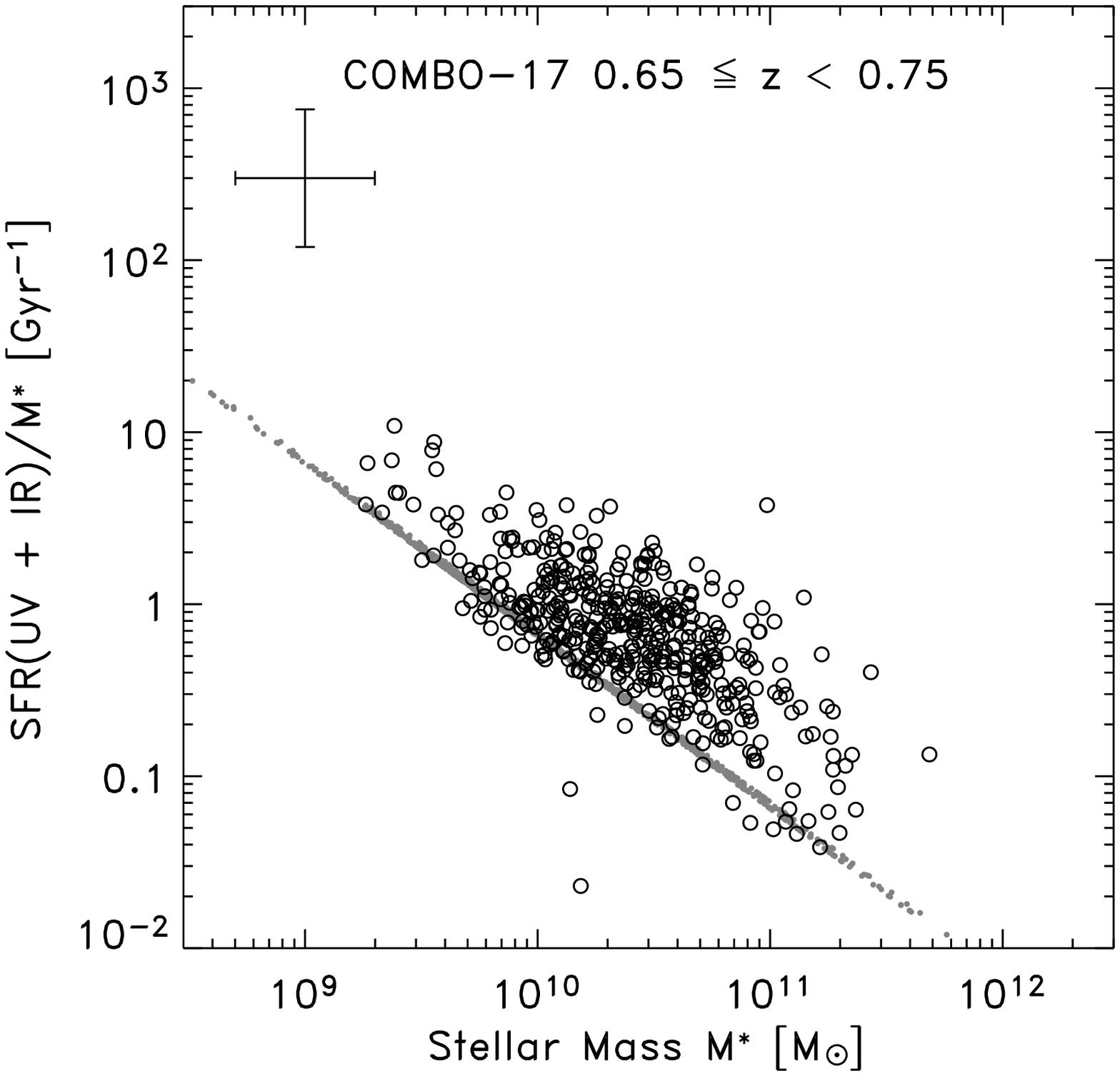}
  \end{minipage}%
}
\vspace{-20pt}
   \caption{The specific SFR as a function of galaxy stellar mass.
   The Left panel shows the DRG and HDF--N galaxies with 1.5
   $\leq$$z$$\leq$ 3, with symbols the same as in figure~1.  The Right
   panel shows galaxies from COMBO--17 (as labeled).  Open circles
   show COMBO--17 galaxies with 24~\micron\ detections, small filled
   symbols show upper limits for galaxies undetected at 24~\micron.}
\vspace{-18pt}
 \end{figure}

Figure~3 shows the specific SFRs ($\Psi/\mcal$)
derived from the masses and SFRs for the DRGs, where the SFRs are
calculated from the summed UV and IR emission.
The figure also shows the specific SFRs for galaxies at lower redshift
from COMBO--17 (Wolf et al.\ 2003), which overlaps with the GTO
\spitzer\ 24~\micron\ imaging.  The SFRs for the COMBO--17 galaxies
are calculated in an analogous manner as for the DRGs.    The massive
DRGs at 1.5$\leq$$z$$\leq$3 have high specific SFRs: DRGs with
$\mcal$$>$$10^{11}$~\msol\ and 1.5$\leq$$z$$\leq$3 have
$\Psi/\mcal$$\sim$0.2--10 Gyr$^{-1}$ (excluding X--ray sources).   In
contrast, at $z$$\lsim$ 0.75 there is an apparent lack of galaxies
with high specific SFRs and high stellar masses: galaxies with
$\mcal$$\geq$$10^{11}$~\msol\ have $\Psi/\mcal$$\sim$0.1--1
Gyr$^{-1}$.

We define the integrated specific SFR as the
ratio of the sum of the SFRs, $\Psi_i$, to the sum of their stellar
masses, $\mcal_i$, $\Upsilon$$\equiv$$\sum_i
\Psi_i / {\sum_i\mcal_i}$,  summed over all $i$ galaxies.  This is
essentially just the ratio of the SFR density to the stellar mass
density for a volume--limited sample of galaxies.    Figure~4 shows
the integrated specific SFRs for DRGs at $z$$\sim$1.5--3.0 and
COMBO--17 at $z\sim 0.4$ and 0.7 with $\mcal \geq 10^{11}$~\msol\ (see
Papovich et al.\ 2006). The data point for the DRGs includes all
objects with $\mcal$$\geq$$10^{11}$~\msol, and assumes that
24~\micron--undetected DRGs have no star formation.   The error--box
lower bound shows what happens if we exclude objects with X--ray
detections or IR colors indicative of AGN.  The error--box upper
bound  shows what happens if we calculate SFRs for the
24~\micron--\textit{undetected} DRGs assuming they have
$f_\nu(24\micron)$=60~\ujy,  the 50\% completeness limit.   The
integrated specific SFR in galaxies with \mcal$>$$10^{11}$~\msol\
declines by more than an order of magnitude from $z$$\sim$1.5--3 to
$z$$\lsim$0.7. Our results indicate that the relative star--formation
in massive galaxies is reduced for $z$$\lsim$1 as galaxies with lower
stellar masses have higher specific SFRs, supporting the so--called
``downsizing'' paradigm.

   \begin{figure}
   \centering
   \includegraphics[width=0.7\linewidth]{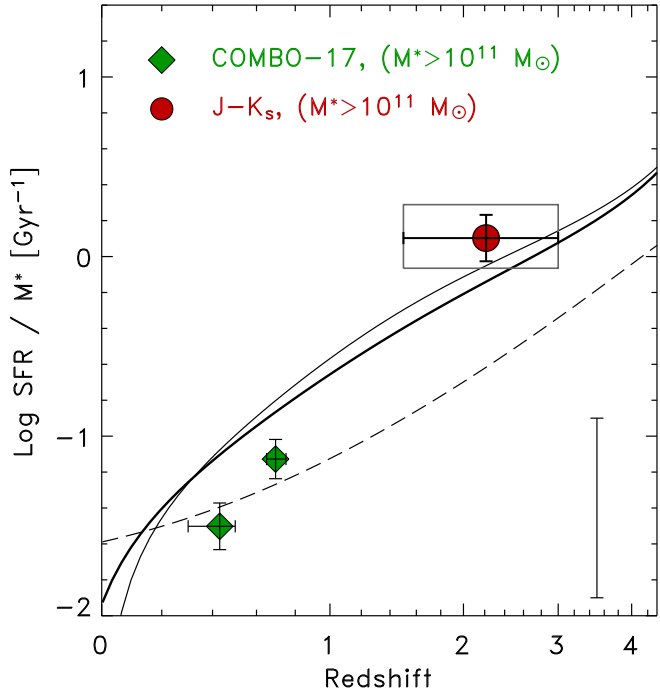}
\vspace{-18pt}
  \caption{Evolution of the integrated specific SFR, the ratio of the
  total SFR to the total stellar mass (from Papovich et al.\ 2006).
  The curves show the expected evolution from the global SFR density
  (solid lines, Cole et al.\ 2001, thick line includes correction for
  dust extinction; dashed line, Hernquist \& Springel 2003).  Data
  points show results for galaxies with $\geq$$10^{11}$ \msol.
  Filled circle corresponds to the DRGs; filled diamonds correspond to
  the COMBO--17 galaxies. The inset error bar shows an estimate on the
  systematics.}
\vspace{-12pt}
   \end{figure}

Figure~4 also shows the specific SFR integrated over all galaxies (not
just the most massive); this is the ratio of the cosmic SFR density
to its integral, $\Upsilon$$=$$\dot{\rho}_\ast /
\int \dot{\rho}_\ast\, dt$.   There is a decrease in the global
specific SFR with decreasing redshift.  The evolution in the
integrated specific SFR in massive galaxies is accelerated relative to
the global value.   Galaxies with $\mcal$$\ge$$10^{11}$~\msol\ were
forming stars at or slightly above the rate integrated over all
galaxies at $z$$\sim$1.5--3.   In contrast, by $z\lsim 1$ galaxies
with $\mcal$ $\ge$$10^{11}$~\msol\ have an integrated specific SFR
much lower than the global value.   \textit{Thus, by $z$$\lsim$1.5
massive galaxies have formed most of their stellar mass, and
lower--mass galaxies dominate the cosmic SFR density} (see also
Papovich et al.\ 2006).

\begin{figure}[h]
\vbox{  \begin{minipage}[t]{0.5\linewidth}
  \centering
  \includegraphics[width=0.95\linewidth]{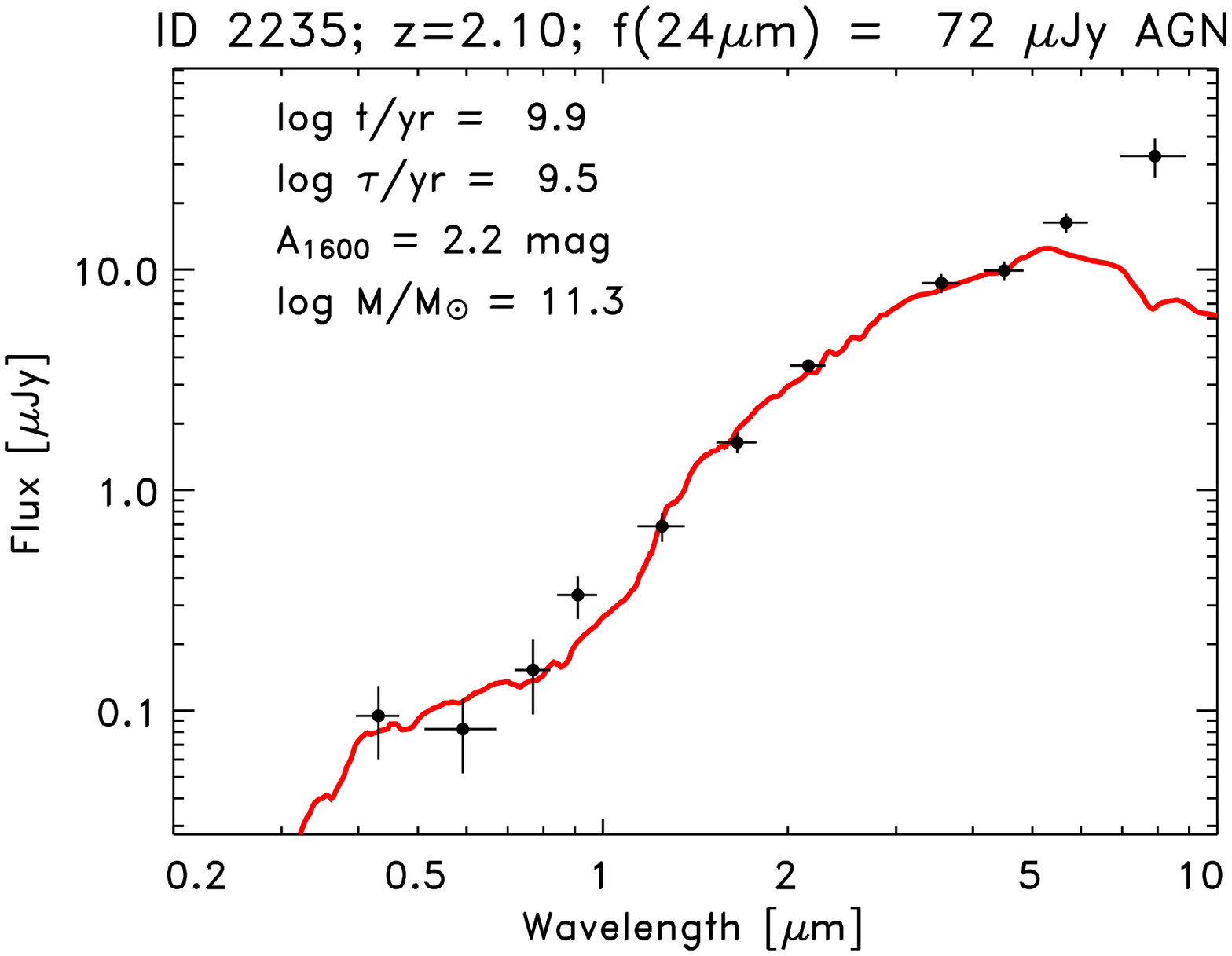}
  \end{minipage}%
  \begin{minipage}[t]{0.5\textwidth}
  \centering
 \includegraphics[width=0.95\linewidth]{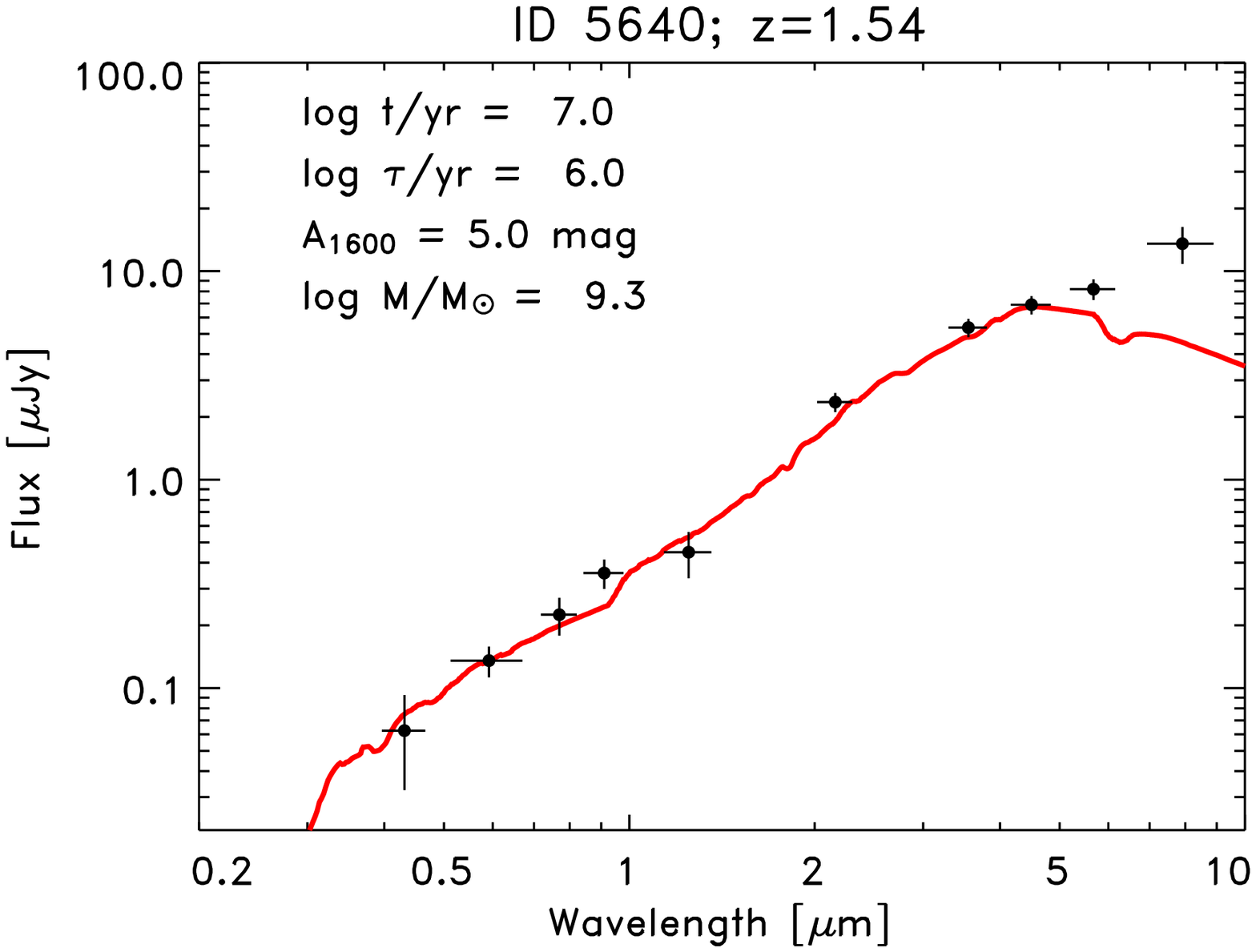}
  \end{minipage}%
}
\vspace{-6pt}
\vbox{  \begin{minipage}[t]{0.5\linewidth}
  \centering
  \includegraphics[width=0.95\linewidth]{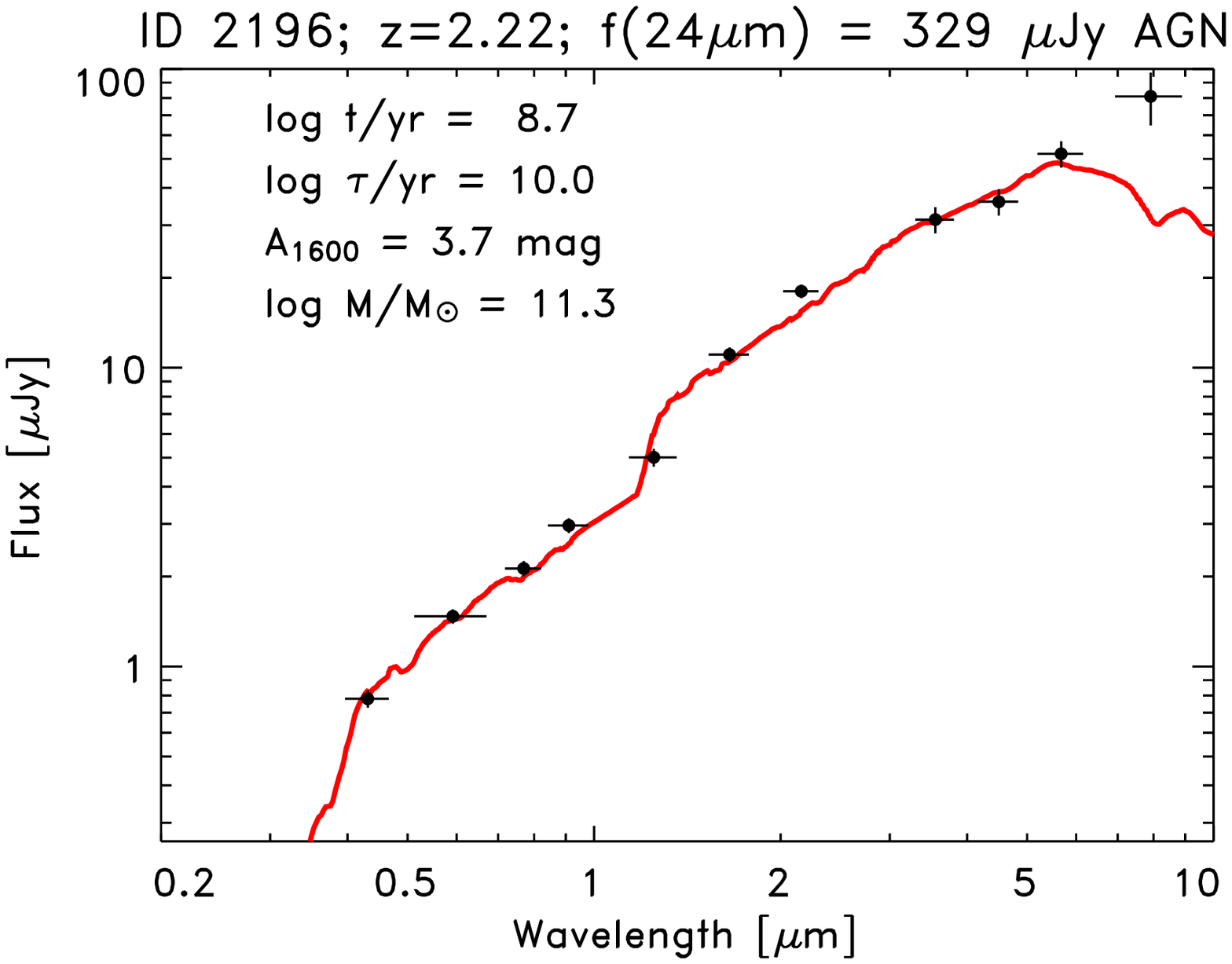}
  \end{minipage}%
  \begin{minipage}[t]{0.5\textwidth}
  \centering
 \includegraphics[width=0.95\linewidth]{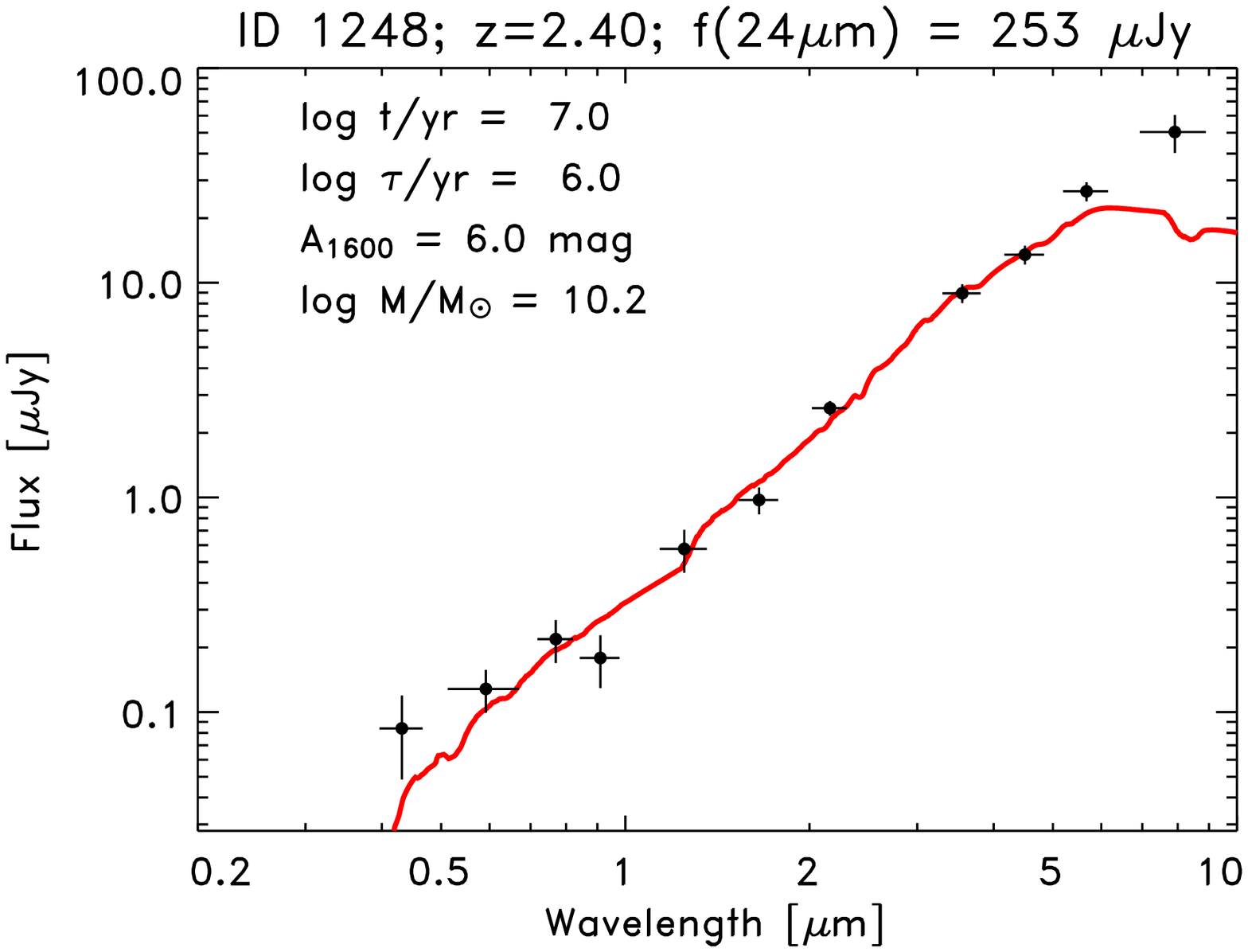}
  \end{minipage}%
}
\vspace{-18pt}
   \caption{SEDs of DRGs with
putative AGN.  The data points in each panel show the rest--frame UV
to near--IR SED from the ACS (0.4--1\micron), ISAAC (1--2.2\micron),
and IRAC (3--8\micron) photometry.  The solid line shows the best--fit
model SED to the ACS, ISAAC, and IRAC \mone\ and \mtwo\ photometry.
The two galaxies in the Left panels have X--ray detections, but the
two in the Right panels do not.  In all cases the galaxies show excess
emission in the IRAC \mthree\ and \mfour\ bands, presumably from hot
dust near the AGN. }
\vspace{-24pt}
 \end{figure}

\section{Uncertainties on Mid-IR--Derived IR Luminosities}

\noindent{\bf 3.1 AGN Contribution to the mid--IR Emission}
\vspace{8pt}

\noindent Many of the massive galaxies at $z$$\sim$1.5--3 are detected in the
deep X--ray data.  In figures~1 and 3, the X-ray--detected DRGs tend
to have the highest inferred IR luminosities and specific SFRs.   Many
of these objects have $\lir$$\geq$$10^{13}$~\lsol, comparable to PG
quasars, which have warm thermal dust temperatures (Haas et al.\
2003).   The X-ray to optical flux ratios in these objects imply the
presence of an AGN with $L_X$$\gsim$$10^{42}$~erg~s$^{-1}$ (for $z$
$\gsim$1.5, see Alexander et al.\ 2003, Papovich et al.\ 2006).   In
addition, many authors are finding AGN candidates based on rest--frame
near--IR colors from \spitzer\ observations.  The X--ray, UV, and
optical emission in these AGN is heavily obscured by gas and dust, and
they are often missed in deep X--ray surveys (e.g., Donley et al.\
2005, Stern et al.\ 2005, Alonso--Herrero et al.\ 2006, Barmby et al.\
2006).   Roughly 25\% of the non--X-ray detected DRGs in the GOODS--S
field satisfy the IRAC color--selection for AGN from Stern et al.\
(2005).  Of these, roughly one--half have ACS--through--IRAC colors
consistent with dust--enshrouded AGN.   Figure~5 shows examples of
these galaxies with and with--out X--ray detections.  In all cases,
the galaxies show a clear flux excess in their IRAC \mthree\ and
\mfour\ photometry, presumably arising from the AGN.  Combined with
the 15\% of DRGs detected in the X--rays, up to 25\% of the DRG
population host AGN. 

If AGN contribute to the observed 24~\micron\ emission in galaxies at
$z$$\sim$1.5--3, then they can affect the inferred IR luminosities.  For
example, although the Chary \& Elbaz (2001) and Dale \& Helou (2002)
IR templates include galxaies with $\lir$$\gsim$ $10^{13}$~\lsol,
using an IR template for Mrk~231 --- with a known AGN and warmer dust
temperature --- would reduce the inferred IR luminosity for galaxies
at $z$$\sim$ 1.5--3 by factors of $\sim$2--5.   To limit the effects
of any bias in the inferred IR luminosities caused by AGN, we
restricted the error box on the integrated specific SFR for the DRGs
in figure~4 to only those galaxies with no X--ray or IR indications
for AGN.

Even when AGN are present, it is unclear whether star--formation or
the AGN activity domninates the bolometric IR luminosity.  Alexander
et al.\ (2005) demonstrated that a high fraction ($\sim$80\%) of
sub--mm galaxies are detected in 2~Msec \textit{Chandra} X--ray data.
However, the X-ray--detected sub--mm galaxies have IR to X--ray
luminosity ratios up to an order of magnitude higher than what is
expected for AGN alone.  This suggests that both star--formation and
AGN contribute to the bolometric emission.  Similarly, Frayer et al.\
(2003) report a near--IR spectrum of a sub--mm galaxy SMM J04431+0210
at $z$$\sim$ 2.5.  This galaxy has $J-\ks$$\simeq$ 3.2~mag, qualifying
as a DRG.   The spectrum shows that the galaxy nucleus has a low \ha\
to [N{\small II}] flux ratio consistent with ionization from an AGN.
However, \ha\ is spatially resolved in their spectrum and the
[N{\small II}] line strength drops off in the off--nucleus spectrum.
This implies extended star--formation beyond the nucleus, which
presumably contributes to the inferred IR luminosity.  Therefore, it
seems that both AGN and star--formation occur simultaneously in
high--redshift IR--detected galaxies and both probably contribute to
the IR emission.

The high AGN occurrence in DRGs and sub--mm galaxies provides some
evidence that massive galaxies at $z$$\sim$1.5--3 simultaneously form
stars and grow supermassive black holes.   Although speculative, the
presence of AGN in these galaxies may provide the impetus for the
present--day relation between black hole  and bulge mass, and/or
provide the feedback necessary to squelch star--formation (e.g.,
Kauffmann et al.\ 2004), moving the galaxies onto the red sequence.
Emission line ratios from high--resolution spectroscopy at near--IR
and/or mid--IR wavelengths may help constrain the ionization state of
the galaxies' gas, identifying the fraction of galaxies with strong
AGN activity as a function of mass at these redshifts.

\vspace{8pt}
\noindent{\bf 3.2 Uncertainties in the Shape of the IR Spectral Energy
Distribution}
\vspace{8pt}

\noindent Although the mid--IR (5--15~\micron) emission broadly
correlates with the total IR luminosity (e.g., Chary \& Elbaz 2001),
there exists considerable scatter.  Partly this arises because
galaxies of a given IR luminosity show a range in the strength of
their mid--IR emission (from polycyclic aromatic hydrocarbons, PAHs)
and absorption features (e.g., Armus et al.\ 2004).  This
variation is reflected in a comparison of various model IR spectral
energy distributions in the literature (e.g., see Le Floc'h et al.\ 2005).
For example, if we used the IR templates of Chary \& Elbaz (2001)
instead of those of Dale \& Helou (2002), then we would derive IR
luminosities a factor of 2--3 higher for $z$$\sim$1.5--3 galaxies with
$\lir$$\sim$$10^{12.5-13}$~\lsol.   In a recent study of galaxies at
$z$$\leq$1.2 detected at \textit{ISO} 15~\micron\ and
\spitzer\ 24~\micron\ in the northern GOODS field, Marcillac et al.\
(2006) found that, compared to Chary \& Elbaz (2001), the IR models of
Dale \& Helou (2002) provide a tighter correlation between \lir\
derived from the mid--IR and radio flux densities with a scatter of
40\%, suggesting the latter may better reflect reality. 

Some scatter is inherent in converting the 24~\micron\ flux
densities to total IR luminosity.  Several studies of the mid--IR
colors of galaxies to $z$$\sim$1 show that their  15 and 16~\micron\
to 24~\micron\ colors have more scatter than predicted by simple
models that map a single IR template to a given IR luminosity (e.g.,
Elbaz et al.\ 2005, Teplitz et al.\ 2005, Marcillac et al.\ 2006).
Chapman et al.\ (2003) find that the temperature--luminosity
distribution in local IR--luminous galaxies has a factor of 2--3
scatter in the IR--luminosity for galaxies of a given dust
temperature.   However, Daddi et al.\ (2005) found that the Chary \&
Elbaz (2001) IR model with $\lir$=$10^{12.2}$~\lsol\ fits the average
spectral energy distribution of 24~\micron--detected $BzK$ objects at
$\langle z \rangle$=1.9 in the northern GOODS field.  Therefore,
\textit{while on average high--redshift galaxies have IR spectral
energy distributions consistent with the models, individually there
remains significant scatter}.

\section{Summary}

In this contribution, I discussed star--formation and AGN activity in
massive galaxies ($\gsim$$10^{11}$~\msol) at $z$$\sim$1--3 using
observations from the \spitzer\ Space Telescope at 3--24~\micron.
Interestingly, the majority ($\gsim$50\%) of these objects have
$f_\nu(24\micron)$$\geq$50~\ujy, which if attributed to star formation
implies SFRs of $\simeq$100--1000~\msol\ yr$^{-1}$.    Galaxies at
$z$$\sim$1.5--3 with $\mcal$$\geq$$10^{11}$~\msol\ have specific SFRs
equal to or exceeding the global average value.  In contrast, galaxies
at $z$$\sim$0.3--0.75 with these masses have specific SFRs less than
the global average, and more than 10$\times$ lower than that at
$z$$\sim$1.5--3.   Therefore, by $z$$\lsim$1.5 massive galaxies have
formed most of their stellar mass, and lower--mass galaxies dominate
the SFR density.

As many as 25\% of the massive galaxies at $z$$\gsim$1.5 host AGN.
The high  AGN occurrence in massive galaxies at $z$$\sim$ 1.5--3
provides evidence that they are simultaneously forming stars and
growing supermassive black holes.   This may provide the impetus for
the present--day black-hole--bulge-mass relation and/or provide the
feedback necessary to squelch star--formation in such galaxies, moving
them onto the red sequence.

The largest source of uncertainty results from systematic errors on
the bolometric corrections between the observed \spitzer\ 24~\micron\
data and total IR luminosity.  While on average high--redshift
galaxies have IR spectral energy distributions consistent with the
models, individually there remains significant scatter.  Future work
is needed to improve our understanding of the distribution between the
mid--IR (rest--frame 5--15~\micron) emission and total IR luminosity
in galaxies at $z$$\sim$1.5--3.  It may be possible to use ``average''
24--to--70 and 160~\micron\ colors of distant galaxies to further
constrain the distribution and scatter of galaxies' IR spectral energy
distributions.  The upcoming \textit{Herschel} Space Observatory (and
eventually \textit{SAFIR}) will mitigate this problem by measuring the
far--IR emission of distant galaxies directly.

\begin{acknowledgements}
I wish to thank the conference organizers for their hard work in
planning an outstanding meeting in an extremely beautiful locale.  I
look forward to the next ``Extreme Starbursts'' meeting.   I am
grateful for my colleagues on the MIPS GTO and GOODS teams for this
research; I am thankful for their continued collaboration. In
particular I am indebted to L.~Moustakas, M.~Dickinson, E.~Le~Floc'h,
E.~Daddi, and G.~Rieke.   I also am grateful for the AAS International
Travel Grant, which made the trip to this meeting possible.   Support
for this work was provided by NASA through the Spitzer Space Telescope
Fellowship Program, through a contract issued by the Jet Propulsion
Laboratory (JPL), California Institute of Technology (Caltech) under a
contract with NASA.

\end{acknowledgements}



\label{lastpage}

\end{document}